\documentstyle{sup2}
\include{psfig}
\vspace{-8pt}

\title[Superlattices and Microstructures, Vol.\ ??, No.\ ??, 1999]
{Multiple Andreev reflections as a transport problem\\
in energy space}

\author[Superlattices and Microstructures, Vol.\ ??, No.\ ??, 1999]
{G\"oran Johansson$^{1}$ and G\"oran Wendin $^{1}$ \cr\vspace{10pt}
{\normalsize\it $^{1}$ Chalmers University of Technology and G\"oteborg University,
S-412 96 G\"oteborg, Sweden}\cr
Kateryna~N.~Bratus'$^{1,2}$ and Vitaly~S.~Shumeiko$^{1,2}$\\
{\normalsize\it $^{2}$ B. Verkin Institute of Low Temperature Physics and Engineering,
310164 Kharkov, Ukraine}\cr
}

\begin{document}
\label{firstpage}
\maketitle
\sloppy
\begin{center}
\received{(Submitted 12 November 1998)}
\end{center}

\begin{abstract}
We present an approach for analyzing the dc current in voltage biased
quantum superconducting junctions. 
By separating terms from different $n$-particle processes,
we find that the $n$-particle
current can be mapped on the problem of wave transport through
a potential structure with $n$ barriers. 
We discuss the relation between resonances in such structures
and the subgap structures in the current-voltage characteristics.
At zero temperature we find, exactly, that only processes creating 
real excitations contribute to the current. Our results are valid for a general
SXS-junction, where the X-region is an arbitrary non-superconducting
region described by an energy-dependent transfer matrix.
\end{abstract}

\section{Introduction}
Multiple Andreev reflection (MAR), first suggested
by Klapwijk, Blonder, and Tinkham \cite{KBT} for explaning the subharmonic 
gap structure (SGS) in 
SNS junctions, is presently accepted as a general mechanism of 
subgap current transport in superconducting junctions. During the last 
five years a considerable effort has gone into developing a 
consistent theory of MAR capable of including effects of
quantum coherence and normal electron scattering by the junction.
Different techniques have been 
used for calculating current-voltage characteristics, including 
various modifications of Keldysh formalism 
\cite{Arn2,Zik,Aminov,Ye1} and the Landauer-B\"uttiker 
scattering method \cite{Kum,Sh1,Sh2,BB,Av,Hurd1}. The 
theory has been extended to resonant tunnel junctions 
\cite{Ye2,Golub,Joh1,Joh2} and junctions of d-wave 
superconductors \cite{Hurd2}. The theory has
successfully explained the SGS in transmissive 
planar junctions \cite{Kle} and atomic-size point contacts \cite{Jan}.
The good agreement between the theory and experimental data 
obtained for one-channel tunnel junctions without fitting 
parameters  \cite{Post} has provided a firm basis for further investigations - 
revealing the transport channels and transmissivity of individual 
channels in single-atom contacts \cite{Urb,Scheer}. 

Despite successful numerical calculations of subgap current-
voltage characteristics for different types of junctions, the general 
understanding of MAR is still not sufficient, and 
interpretation of particular features  of SGS, e.g. current peaks, 
presents difficulties. This especially concerns resonant MAR, where  
the energy dispersion of electron scattering phase shifts is important.

In this paper, we present an approach where MAR is
treated as a wave propagation problem in energy space (cf. \cite{BB}). 
Using scattering theory formalism, we derive an adequate 
mapping of MAR onto a transmission problem through a 1D wave guide with a 
built in multiple tunnel barrier structure. All parameters of this structure 
are uniquely determined by the junction characteristics (normal and Andreev 
scattering amplitudes) and by the applied voltage. 
In terms of such a mapping, the onsets and peaks of SGS are explained 
in terms of resonances in the transmission along the energy axis.
The mapping allows us consistently to separate currents 
associated with $n$-particle transmission processes 
\cite{SW} and to prove the cancellation of non-physical ground 
state currents, which is equivalent to the Pauli exclusion principle.

\section{Model and ansatz}
We consider two superconducting reservoirs connected to a single
normal conducting channel which may contain
tunnel barriers (X-region) (see Fig. 1).
The transport properties of this channel
is described by its transfer matrix.
Using an energy dependent transfer matrix we can model any
effective single-particle potential structure, including
the important scattering phase shifts in long and resonant junctions.
In general the transfer matrix is also voltage dependent,
and deriving this dependence for any given physical model is a problem
in its own right. But having found this transfer matrix, we
can determine the current when the contacting reservoirs
are in the superconducting state. 

We place the origin of our
coordinate system in the middle of the junction; the
normal-superconducting (NS) interfaces are then located at
$-L/2$ and $L/2$, where $L$ is the length of the junction
(Fig. 1). Due to the point-contact geometry of our junction,
the superconducting pair potential can be considered 
steplike, $\Delta(x)=\Delta_{0}\Theta(|x|-L/2)$.

We will construct scattering
states in terms of the transfer matrix of the X-region.
To this end we introduce two auxilliary normal regions,
between the X-region and the superconducting electrodes.
These regions are assumed to have the same material parameters
as the electrodes, providing perfect NS-interfaces, and their
length is much smaller than the coherence length.

\begin{figure}
\centerline{\psfig{figure=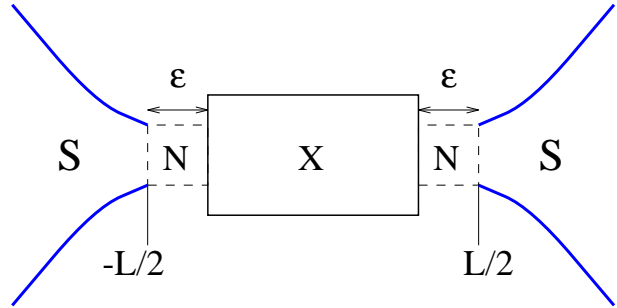,height=4cm}}
\caption{\small Sketch of the junction geometry:
The junction consists of an arbitrary normally conducting,
single channel region (X), connected to two superconducting
reservoirs (S). Two ballistic normal regions (N)
of length $\epsilon$ between the X and S regions are introduced for
convenience. $\epsilon$ is much smaller than the coherence length.}
\end{figure}

To calculate the scattering states we first make an
ansatz in terms of a sum of plane-wave solutions to the
Bogoliubov-de Gennes (BdG) equation in the left/right
auxiliary normal regions ($\Psi_{LN}/\Psi_{RN}$) and in
the left/right superconductors ($\Psi_{LS}/\Psi_{RS}$).
In superconductors, the energy is commonly counted from the chemical
potential which is motivated by the electron-hole symmetry.
This is not convenient in voltage biased
junctions because the chemical potentials in the left ($\mu_{LS}$)
and right ($\mu_{RS}$) electrodes are different.
A global reference of energy, which is particularly desirable in the case
of an energy dependent transfer matrix, can be introduced by making
different gauge transformations in the left and right
superconducting electrodes. This procedure gives rise to
the appearance of time-dependent phase factors in the
wave functions of the superconducting electrodes implying inelastic
scattering.
To get symmetrical expressions for quasiparticles incoming from
the left and from the right we choose $(\mu_{LS}+\mu_{RS})/2$ as a global 
reference of energy, i.e. the reference energy is in
the middle of the chemical potentials in the left and
right superconductors. We thus get different time-dependences
for electron and hole components in \em both \rm superconductors,

\begin{eqnarray}
\label{LNansatz}
\Psi_{LN} &=& \sum_{n=\bar{\infty}}^{\infty} \left( \begin{array}{c}
\alpha_{n} e^{i k^{e}_{n} (x+L/2)} + \beta_{n} e^{-i k^{e}_{n} (x+L/2)} \\
\gamma_{n} e^{i k^{h}_{n} (x+L/2)} + \delta_{n} e^{-i k^{h}_{n} (x+L/2)} 
\end{array} \right) e^{-i(E_{n}+eV/2) t} , \\
\label{RNansatz}
\Psi_{RN} &=& \sum_{n=\bar{\infty}}^{\infty} \left( \begin{array}{c}
\alpha_{n}' e^{i k^{e}_{n} (x-L/2)} + \beta_{n}' e^{-i k^{e}_{n} (x-L/2)} \\
\gamma_{n}' e^{i k^{h}_{n} (x-L/2)} + \delta_{n}' e^{-i k^{h}_{n} (x-L/2)} 
\end{array} \right) e^{-i(E_{n}+eV/2) t} , \\
\label{LSansatz}  
\Psi_{LS} &=& e^{-i\sigma_{z}eV/2t}\cdot
\left[\left(\delta_{1\sigma}\hat{u}^{e}_{n}e^{+i\tilde{k}^{e}_{0}(x+L/2)} +
\delta_{2\sigma} \hat{u}_{n}^{h}e^{-i \tilde{k}^{h}_{0} (x+L/2)} \right)
e^{-iE t} + \right. \nonumber\\
&+& \left. \sum_{n=\bar{\infty}}^{\infty}
\left(A_{n} \hat{u}_{n}^{e}e^{-i \tilde{k}^{e}_{n} (x+L/2)} +
B_{n} \hat{u}_{n}^{h} e^{+i \tilde{k}^{h}_{n} (x+L/2)} \right)
e^{-iE_{n} t} \right] , \\
\Psi_{RS} &=& e^{i\sigma_{z}eV/2t}\cdot 
\left[\left(\delta_{3\sigma}\hat{u}_{n}^{e} e^{-i\tilde{k}^{e}_{0}(x-L/2)} +
\delta_{4\sigma} \hat{u}_{n}^{h}  e^{+i \tilde{k}^{h}_{0} (x-L/2)} \right)
e^{-iE t} + \right. \nonumber \\
&+& \left. \sum_{n=\bar{\infty}}^{\infty}
\left(C_{n} \hat{u}_{n}^{e}e^{i \tilde{k}^{e}_{n} (x-L/2)} +
F_{n} \hat{u}_{n}^{h}e^{-i\tilde{k}^{h}_{n}(x-L/2)}\right)
e^{-iE_{n} t} \right].
\label{RSansatz}  
\end{eqnarray}

In the ansatz in Eqs. (\ref{LNansatz})-(\ref{RSansatz}),
$E$ denotes the energy of the incoming particle,
$E_{n} = E + neV$, and the index $\sigma \in \{1,2,3,4\}$ denotes
the incoming quasiparticles $\{e^{\rightarrow},h^{\rightarrow}
,e^{\leftarrow},h^{\leftarrow}\}$ respectively.
$\hat{u}_{n}^{e}=(u_{n},v_{n})$ and $\hat{u}_{n}^{h}=(v_{n},u_{n})$
are vectors describing bulk electron- and hole-like
quasiparticles respectively.

By matching the wave functions across the 
NS intefaces we get the boundary conditions for the wave function
in the normal region, including the source term. By defining
coefficient vectors $\hat{\alpha}_{n}=(\alpha_{n},\beta_{n})$
and $\hat{\gamma}_{n}=(\gamma_{n},\delta_{n})$, and similar
for the primed quantities in Eq.~(\ref{RNansatz}), we may write these
boundary conditions in the following form,


\begin{equation}
\label{NSeq}
\hat{\alpha}_{n} = {\bf U}_{n} \hat{\gamma}_{n-1} +
\delta_{n0} (u_{0}^{2}-v_{0}^{2}) \left( \begin{array}{c}
\delta_{1\sigma}/u_{0} \\
-\delta_{2\sigma}/v_{0} \end{array} \right),
\hspace{.5cm}
\hat{\gamma}'_{n} = {\bf U}_{n} \hat{\alpha}'_{n-1} +
\delta_{n0} (u_{0}^{2}-v_{0}^{2}) \left( \begin{array}{c}
\delta_{4\sigma}/u_{0} \\
-\delta_{3\sigma}/v_{0} \end{array} \right) ,
\end{equation}
where ${\bf U}_{n}=diag(a_{n},a_{n}^{-1})$ is the Andreev
reflection matrix and $a_{n}$ is the amplitude of Andreev reflection,

\begin{equation}
a_{n} = \frac{v_{n}}{u_{n}} = \left\{ \begin{array}{cc}
\left(E_{n}-\mbox{sign}(E_{n})\sqrt{E_{n}^{2}-\Delta^{2}}\right)
/\Delta & |E| > \Delta \\
\left(E_{n}-i\sqrt{\Delta^{2}-E_{n}^{2}}\right)
/\Delta & |E| < \Delta
\end{array} \right. .
\end{equation}
In this derivation we have neglected the difference in wave vectors
between the normal and superconducting regions
(quasiclassical approximation). The coefficients
in Eqs. (\ref{LNansatz})-(\ref{RNansatz}) are connected
by the transfer matrix ${\bf T}(E)$ describing the X-region,

\begin{equation}
\label{TMeq}
\hat{\alpha}_{n} = {\bf T}(E_{n}+eV/2) \hat{\alpha}'_{n}
\hspace{1cm}
\hat{\gamma}_{n} = {\bf T}(-(E_{n}+eV/2)) \hat{\gamma}'_{n} .
\end{equation}

\noindent Equations (\ref{NSeq})-(\ref{TMeq}), together
with the boundary condition of vanishing coefficients for
$|n| \rightarrow \infty$, completely determine the scattering states.

Checking equations (\ref{NSeq})-(\ref{TMeq}) one may find that in
any scattering state half of the coefficients in the ansatz
will be zero. For example, for $\sigma \in \{1,2\}$ the nonzero
coefficients will be $\hat{\alpha}_{2m}$ and
$\hat{\gamma}_{2m-1}$, while for $\sigma \in \{3,4\}$ 
$\hat{\alpha}_{2m-1}$ and $\hat{\gamma}_{2m}$ will be nonzero.
To get rid of the redundant coefficients
we define new coefficients $c_{n\pm}$. For $\sigma \in \{1,2\}$
the definition reads $\hat{c}_{2m+} = \hat{\alpha}_{2m}$,
$\hat{c}_{2m-}  =  \hat{\gamma}_{2m-1}$,
$\hat{c}_{(2m+1)+}  =  \hat{\gamma}'_{2m+1}$ and
$\hat{c}_{(2m+1)-}  =  \hat{\alpha}'_{2m}$.
These coefficients are connected by transfer matrices ${\bf T}^{l}_{n}$ 
with even indices ${\bf T}^{l}_{2m} = \left[{\bf T}(E_{2m}+eV/2)\right]^{-1}$
and odd indices ${\bf T}^{l}_{2m-1} = {\bf T}(-(E_{2m-1}+eV/2))$.
For the purpose of
treating quasiparticles incoming from the left and right on an equal
footing we introduce similar notations for $\sigma \in \{3,4\}$:
$\hat{c}_{2m+}  =  \hat{\gamma}'_{2m}$,  
$\hat{c}_{2m-}  =  \hat{\alpha}'_{2m-1}$,
$\hat{c}_{(2m+1)+}  =  \hat{\alpha}_{2m+1}$ and
$\hat{c}_{(2m+1)-}  =  \hat{\gamma}_{2m}$ and matrices 
${\bf T}^{r}_{2m} = {\bf T}(-(E_{2m}+eV/2))$ and
${\bf T}^{r}_{2m-1} = \left[{\bf T}(E_{2m-1}+eV/2)\right]^{-1}$.

Using these definitions the equations (\ref{NSeq})-(\ref{TMeq})
can, for quasiparticles incoming from both left and right, be written as:

\begin{equation}
\label{newRec}
\hat{c}_{n+} = {\bf U}_{n} \hat{c}_{n-} +
\delta_{n0} (u_{0}^{2}-v_{0}^{2}) \left( \begin{array}{c}
(\delta_{1\sigma}+\delta_{4\sigma})/u_{0} \\
-(\delta_{2\sigma}+\delta_{3\sigma})/v_{0} \end{array} \right),
\hspace{1cm}
\hat{c}_{(n+1)-} = {\bf T}_{n} \hat{c}_{n+}.
\end{equation}

\noindent Note that even though these equations
are formally identical for quasiparticles incoming from both
left and right, the ${\bf T}_{n}$-matrices involved in the
two cases are different.
In Figure 2a we show the structure of
Eqs.~(\ref{newRec}) for an electron-like quasiparticle
incoming from the left. 

\begin{figure}
\centerline{\psfig{figure=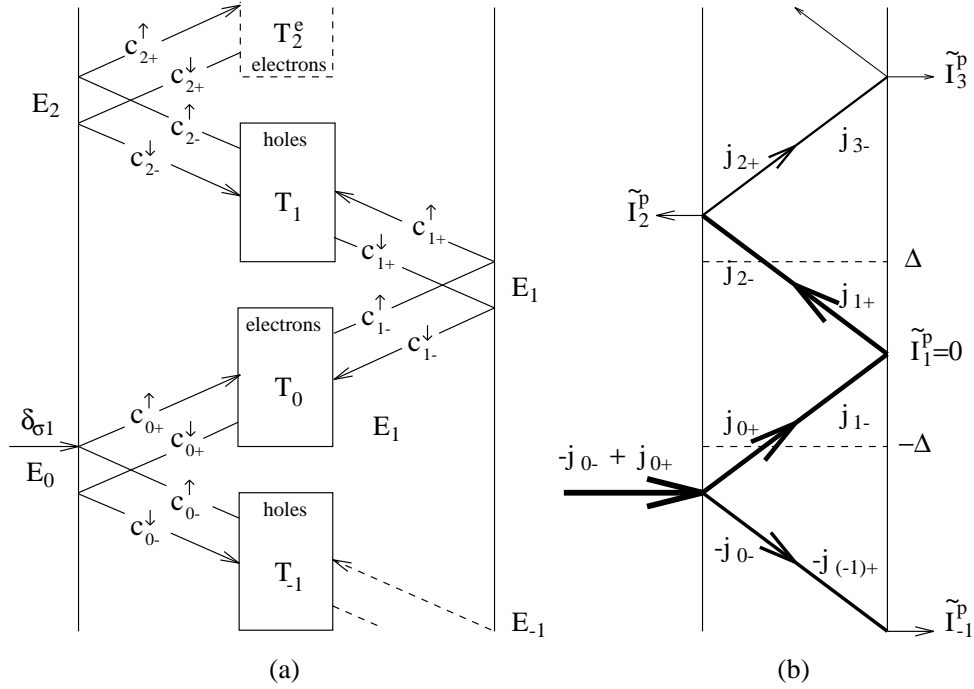,height=9cm}}
\caption{\small Schematic structure of a scattering state (a) and
its probability currents (b). Figure (a) shows schematically
the connection between different coefficients and transfer
matrices for an electronlike quasiparticles incoming from the left
($\sigma=1$). Figure (b) shows schematically how the
probability current from the incoming quasiparticle enters
the normal region and divides into upgoing ($j_{0+}$)
and downgoing ($j_{0-}$) currents. This probability current
then leaks out due to incomplete Andreev reflections outside the gap
($\tilde{I}^{p}_{-1}, \tilde{I}^{p}_{2}$
and $\tilde{I}^{p}_{3}$).}
\end{figure}

\section{Mapping}

Now we are prepared to change focus from
transport in real space to transport in energy space.
The vectors $\hat{c}_{n\pm}$ have been defined so that the upper
element is the coefficient for particles with positive $k$
and the lower element is the coefficient for negative $k$.
We notice that both electrons and holes with positive
$k$-value gain energy in passing the junction (for our choice of the sign
of bias), while particles with negative $k$-value lose energy.
Therefore the upper vector component describes upward motion in
energy space, while the lower component describes downward motion.
To emphasize this motion in energy space we introduce the following
notation for the components of $\hat{c}_{n\pm}$:

\begin{equation}
\hat{c}_{n\pm} = \left( \begin{array}{c}
c^{\uparrow}_{n\pm} \\
c^{\downarrow}_{n\pm} \end{array} \right),
\end{equation}
where $c^{\uparrow}$ indicates upward motion in energy space
and $c^{\downarrow}$ indicates downward motion (see Fig. 2a).
With this notation, Eqs.~(\ref{newRec}) have obvious similarities
to the  problem of wave propagation through a one-dimensional
multibarrier structure. Indeed the coefficient vectors
are connected by simple matrix multiplication
through source-free regions, $n>m \geq 0$ or $0\geq n > m$,

\begin{equation}
\label{hommat}
\hat{c}_{n-}={\bf M}_{nm} \hat{c}_{m+}, \;\;\; \hspace{.3cm}
{\bf M}_{nm} = {\bf T}_{n-1} {\bf U}_{n-1} {\bf T}_{n-2} ...
{\bf T}_{m+1} {\bf U}_{m+1} {\bf T}_{m}.
\end{equation}
The ${\bf M}$-matrices describe a 1D multibarrier structure, where the
${\bf T_{n}}$-matrices provide tunnel barriers, and the ${\bf U}_{n}$-matrices
introduce extra spacing, i.e. phase gained, between these barriers.
Within the superconducting energy gap these ${\bf M}$-matrices have the
properties of ordinary real space Schr\"odinger transfer matrices,
which conserve current:
$det({\bf M})=1$, $\sigma_{z}{\bf M}\sigma_{z}={\bf M}^{\dagger}$.
In our case, the conserved quantity is, $j_{n\pm}=
|c_{n\pm}^{\uparrow}|^{2}-|c_{n\pm}^{\downarrow}|^{2}$, which
can be interpreted as the probability current flowing upwards in
energy space. The probability current $j_{n\pm}$ is related to electron 
and hole currents
in the original real space problem. For example, for quasiparticles 
incoming from the left
$j_{(2n+1)+}=j_{(2n+2)-}$ is probability current
of electrons, while $-j_{2n+}=-j_{(2n+1)-}$ is probability current
of holes (the equalities are provided by the current conserving properties 
of the normal transfer
matrices $T$). The quasiparticle probability current $I^{p}_{n}$,
which is conserved by the BdG equation, consists of the sum of
electron and hole probability currents,
$I^{p}_{n}=j_{(2n+1)+}+(-j_{(2n+1)-})$. 
The probability current $I^{p}_{n}$ is obviously equal to zero inside the
energy gap because of the conservation of the current $j_{n}$. 

The ${\bf M}$-matrices are associated with scattering matrices ${\bf S}$,

\begin{equation}
\left(\begin{array}{c}
c_{m+}^{\downarrow} \\
c_{n-}^{\uparrow} \end{array} \right) =
{\bf S}_{nm} \left(\begin{array}{c}
c_{m+}^{\uparrow} \\
c_{n-}^{\downarrow} \end{array} \right), \hspace{.5cm}
{\bf S}_{nm} = \left(\begin{array}{cc}
r_{nm}^{+} & t_{nm} \\
t_{nm} & r_{nm}^{-} \end{array} \right)
\end{equation}

\noindent where the reflection amplitudes $r^{+}=-M_{21}/M_{22},
r^{-}=M_{12}/M_{22}$ and transmission amplitude $t=1/M_{22}$
satisfy the current conservation condition $|r|^2+|t|^2=1$ within the
superconducting energy gap.
Outside the gap this condition changes into
$|r_{nm}^{\pm}|^2+|t_{nm}|^2 \prod_{p=m+1}^{n-1}|a_{p}|^{-2}\leq 1$,
where the equality holds only for fully transparent junctions.
Since the probability of Andreev reflection is smaller than unity outside
the gap, $|a_{n}|^{2}< 1$ for $|E_{n}|>\Delta$, the current $j_{n}$ is
not conserved, which is related to leakage outside the normal region
due to incomplete Andreev reflection. This leakage is characterized by the
the probability current $I^{p}_{n} \neq 0$. 
The sign of $I^{p}_{n}$ in the above definition was chosen so
that probability current flowing away from the junction is positive, and
since the only incoming current is from the source, $I^{p}_{n} \geq 0$ for
$n\neq 0$.
We will see in the next section that the
dc current can be expressed in terms of this ``leakage''.

\begin{figure}
\centerline{\psfig{figure=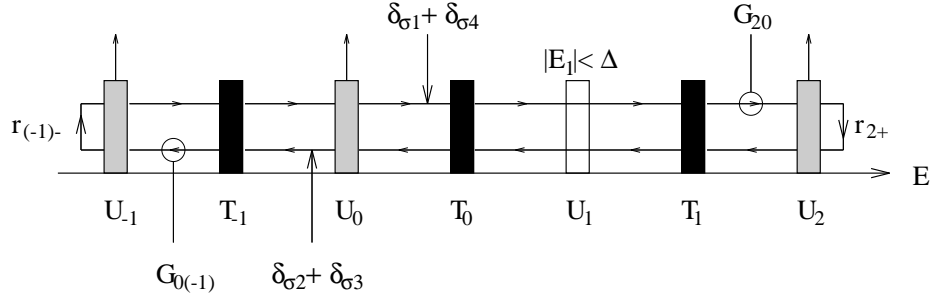,height=4cm}}
\caption{\small Schematic structure of the mapping
of the scattering state equations on a leaky one-dimensional
Schr\"odinger potential. The ${\bf T}_{n}$-matrices introduce
tunnel barriers (black boxes). Inside the energy gap, $|E_{n}|<\Delta$,
the ${\bf U}_{n}$-matrices correspond to extra spacing (white box),
while outside the gap they introduce leaking of particles out of
the channel. The sources ($\delta_{\sigma 1}+\delta_{\sigma 4}$
and $\delta_{\sigma 1}+\delta_{\sigma 4}$) introduce particles into the
potential structure and the amplitude for particles going away from 
the source are $G_{n0}$ and $G_{0(-n)}$, $n>0$. The particles are 
reflected from $\pm\infty$ with reflection amplitudes $r_{n\pm}$.
}
\label{potstructfig}
\end{figure}

Particles are introduced into this ``leaking'' potential 
structure by the source. 
It follows from Eq.~(\ref{newRec}) that the injection of
quasiparticles with positive $k$ ($\sigma\in\{1,4\}$)
introduces upgoing particles directly above $U_{0}$, while
quasiparticles with negative $k$ ($\sigma\in\{2,3\}$) inject downgoing
particles just below $U_{0}$ (see Fig.~\ref{potstructfig}).

\section{Formal solution}
To solve Eqs. (\ref{newRec}) one has to find
two independent homogenous solutions that go to zero for $n=\pm\infty$
respectively, and then to match them at $n=0$. The boundary conditions,

\begin{equation}
\lim_{n\rightarrow\infty} {\bf M}_{nm} \hat{c}_{m+}
=\mbox{\boldmath$0$} \hspace{.3cm} (m\geq 0) ,
\hspace{1cm} \lim_{m\rightarrow -\infty} ({\bf M}_{nm})^{-1} \hat{c}_{n-}
=\mbox{\boldmath$0$} \hspace{.3cm} (n\leq 0) ,
\end{equation}

\noindent fix the ratios $ c^{\downarrow}_{n+}/c^{\uparrow}_{n+}=r_{n+}=
\lim_{m\rightarrow\infty}r^{+}_{mn}$ for $n\geq 0$ and 
$ c^{\uparrow}_{n-}/c^{\downarrow}_{n-}=r_{n-}=
\lim_{m\rightarrow -\infty}r^{-}_{nm}$ for $n\leq 0$.
The quantities $r_{n\pm}$ have the meaning of reflection amplitudes
from $\pm\infty$ for upgoing and downgoing particles respectively.

We can now express the scattering state coefficients
in terms of the matrices $S_{nm}$
and the reflection coefficients $r_{n\pm}$. We choose to
give expressions for $\hat{c}_{n-}$ for $n>0$ and
$\hat{c}_{n+}$ for $n<0$, knowing that
Eq.~(\ref{newRec}) connects $\hat{c}_{n+}$ to $\hat{c}_{n-}$.


\begin{eqnarray}
\label{leftcoff}
\hat{c}_{n-}=
\frac{u_{0}^{2}-v_{0}^{2}}{u_{0}} \cdot
\left[\delta_{1\sigma}+\delta_{2\sigma}a_{0}r^{l}_{0-}\right]
\cdot G^{l}_{n0} \cdot \left( \begin{array}{c}
1 \\
a_{n}^{2}r^{l}_{n+} \end{array} \right) \hspace{.3cm}(n>0) , \\
\hat{c}_{n+}=\frac{u_{0}^{2}-v_{0}^{2}}{u_{0}} \cdot
\left[\delta_{2\sigma}+\delta_{1\sigma}a_{0}r^{l}_{0+}\right]
\cdot G^{l}_{0n} \cdot \left( \begin{array}{c}
a_{n}^{2}r^{l}_{n-} \\
1 \end{array} \right) \hspace{.3cm}(n<0) .
\end{eqnarray}

\noindent The expressions for $\sigma\in\{3,4\}$ is found
from Eq.~($\ref{leftcoff}$) by the substitutions $l\rightarrow r$,
$\delta_{1\sigma}\rightarrow\delta_{4\sigma}$ and
$\delta_{2\sigma}\rightarrow\delta_{3\sigma}$. The expression
for the quantity $G$ in the above equation reads 

\begin{equation}
\label{Gdef}
G_{nm} = \frac{t_{nm}}
{(1-a_{m}^{2}r_{m-}r_{nm}^{+})(1-a_{n}^{2}r_{n+}r_{nm}^{-})-
t_{nm}^{2}a_{m}^{2}a_{n}^{2}r_{m-}r_{n+}}
\hspace{1cm} (n>m) .
\end{equation}

\noindent $G_{nm}$ is the dressed propagator, through
source free regions of the multibarrier structure from point $m$ to point $n$
(see Fig.~\ref{potstructfig}). The difference between
the dressed propagator $G_{nm}$ and the bare transmission
amplitude $t_{nm}$ is due to reflections from the region outside
$[E_{m},E_{n}]$.
Note that $G_{nm}$ in Eq.~(\ref{Gdef}) is expressed through
$r_{n+}$ and $r_{m-}$ (using the fact that the reflection amplitudes from
the same sign of infinity are related by Eq.~(\ref{hommat})). This form is 
vital for the cancellation theorem below (Eq.~(\ref{canctheo})). 

\section{dc Current}

The total dc current through the junction has the form,

\begin{equation}
I_{dc}=\frac{e}{h}
\int_{|E|>\Delta} dE
\frac{|E|}{\sqrt{E^2-\Delta^2}}  f(E) \sum_{\sigma} j^{\sigma}(E) ,
\end{equation}
where the function $f(E)$ denotes the equilibrium population
of the scattering states, originating from bulk electrodes.
The current density $j^{\sigma}(E)$ has the form,

\begin{equation}
j^{\sigma}(E) = \sum_{n=-\infty}^{\infty}
\left(
|\alpha^{\sigma}_{n}(E)|^{2}-|\beta^{\sigma}_{n}(E)|^{2} +
|\gamma^{\sigma}_{n}(E)|^{2}-|\delta^{\sigma}_{n}(E)|^{2}
\right).
\end{equation}

\noindent We can rewrite this in
terms of the current flowing upwards in energy-space, $j_{n\pm}$, 

\begin{equation}
\label{currdens}
j^{\sigma}(E) =\sum_{n=0}^{\infty} j^{\sigma}_{n+}(E) +
\sum_{n=-\infty}^{0} j^{\sigma}_{n-}(E) .
\end{equation}

\noindent We note that for positive $n$ the current $j_{n\pm}$
is positive, while for negative $n$ it is negative, and therefore
the two terms in the above equations have opposite signs.
Using the boundary condition of vanishing coefficients for large $|n|$,
we can rewrite the summation over $j_{n\pm}$ in Eq.~(\ref{currdens})
as a summation over probability currents $I^{p}_{n}(E)$, 

\begin{equation}
j ^{\sigma}(E)=\sum_{n\neq 0} n I^{p\sigma}_{n}(E) .
\label{dc_curr_scatt_state}
\end{equation}
We have now managed to separate the dc current
into a weighted sum over probability currents transmitted
from $E_{0}$ to $E_{n}$, where the weight $n$, is
the number of times the probability current passes the junction.
Since each Andreev reflection at energies within the
superconducting gap is associated with the transfer of charge
$2e$, Eq.~(\ref{dc_curr_scatt_state}) presents the total
charge current as a sum over $n$-particle
transmission processes \cite{SW}.

The total probability
current that flows between $E$ and $E_{n}$, including the 
superconducting density of states, is

\begin{equation}
\label{Ipdef} 
I^{p}_{n}(E)=\frac{|E|}{\sqrt{E^2-\Delta^2}}
\sum_{\sigma}I_{n}^{p\sigma}(E) =
\displaystyle \sum_{\alpha\in\{l,r\}}
(1-|a_{0}|^{2})(1-|a_{n}|^{2})
|G^{\alpha}_{n0}|^{2}(1+|a_{n}r^{\alpha}_{n+}|^{2})
(1+|a_{0}r^{\alpha}_{0-}|^{2}) .
\end{equation}

\noindent This current has no divergences, moreover by using the 
conservation law for probability current one finds that
$I^{p}_{n}(E) \leq 4$. The form of the current in Eq.~(\ref{Ipdef}) can be
interpreted in the following way. The current is proportional to the
probability of transmission from $E$ to $E_n$, and the terms
$(1-|a_{0}|^{2})$ and $(1-|a_{n}|^{2})$ are the probabilities
to enter and leave the normal region respectively. Transmission from $E_{}$
to $E_{n}$ consists of four alternative paths: direct
transmission with the probability $|G_{n0}|^{2}$, transmission with
excursion to and reflection from either minus infinity or plus infinity, 
which yields an
additional factor $|a_{0}r_{0-}|^{2}$ or $|a_{n}r_{n+}|^{2}$ respectively,
and finally transmission with excursions to and reflections from both minus 
infinity and plus infinity.

The probability current $I^{p}_{n}(E)$ in Eq.~(\ref{Ipdef})
obeys the important equation

\begin{equation}
I^{p}_{-n}(E)=I^{p}_{n}(E-neV) ,
\end{equation}

\noindent which implies a detailed balance between probability currents
flowing upwards and downwards in energy space respectively. This balance is 
straightforward to prove
using the equality between $S_{nm}$ for quasiparticles incoming at $E+qeV$
and $S_{(n+q)(m+q)}$ for quasiparticles incoming at $E$.
Using this detailed-balance equation one is able to prove the
exact cancellation, at $T=0$, of currents which do not pass the energy gap,

\begin{equation}
\label{canctheo}
\int_{-\infty}^{-\Delta-neV} I^{p}_{n}(E) dE -
\int_{-\infty}^{-\Delta} I^{p}_{-n}(E) dE = 0 \hspace{1cm} (n>0).
\end{equation}
This cancellation theorem proves the consistency of the single
particle scattering approach to superconducting junctions.
It is well known that within the Landauer-B\"{u}ttiker
scattering approach to elastic tunneling,
e.g. in normal junctions \cite{Lan,But}, the Pauli exclusion principle is
automatically valid due to the exact cancellation of
currents that do not produce real excitations.
The cancellation theorem Eq.~(\ref{canctheo}) extends this
property to inelastic scattering (MAR) in superconducting junctions.

Using Eq.~(\ref{canctheo}), and also the electron-hole symmetry
of the BdG equation ($\sum_{\sigma} j^{\sigma}(-E)=
-\sum_{\sigma} j^{\sigma}(E)$), we arrive at the final formula
for the dc current at finite temperature,

\begin{equation}
I_{dc} = \sum_{n>0} n \left(
\Theta(neV-2\Delta) \int_{\Delta-neV}^{-\Delta} dE
[2f(E)-1]I^{p}_{n}(E) +
2 \int_{-\infty}^{-\Delta-neV} dE
[f(E)-f(E_{n})] I^{p}_{n}(E) \right) .
\label{currformula}
\end{equation}
The obtained formula for the dc current in Eq.~(\ref{currformula})
has only positive terms and consists of two parts. The first part
is related to the creation of real excitations, due to
quasiparticles traversing the energy gap, and it is responsible
for the SGS. This current exists at zero
temperature and decreases with increasing temperature.
The second part is a smooth background current from thermal excitations,
and is exponentially small at low temperatures.




\section{Subgap structure}

Let us analyze SGS using Eqs.~(\ref{Gdef}), (\ref{Ipdef}) and
(\ref{currformula}). The magnitude of the $n$-particle current
$I^{p}_{n}(E)$ is proportional to the transmittivity $|t_{n0}|^2$
of the $n$-barrier tunnel structure, which is estimated as
$|t_{n0}|^2 \propto D^n$ in the absence of resonances,
where $D$ is the transmissivity of the normal junction. 
Therefore, the current in Eq.~(\ref{currformula}) consists
of a step-like structure with the onsets at $eV=2\Delta/n$.
This step-like structure is particularly pronounced in
junctions with low transmittivity $D \ll 1$ \cite{Sh1},
and it is washed out in transparent junctions with $D\approx 1$.
However, even in the latter case, the current decreases exponentially
at low voltages, $I_{dc}\propto D^{2\Delta/eV}$, as soon as $D\neq 1$
\cite{BB}.

This simple step-like structure is complicated by resonances in
$I^{p}_{n}(E)$. One may distinguish three different sources
of resonant behaviour. First there may be normal electron
resonances in the transfer matrices $T_{n}$ \cite{Ye2, Joh1, Joh2}.

Secondly, even in the absence of normal electron resonances
there are specific superconducting resonances in the \em bare \rm
transmission amplitudes $t_{n0}$ due to electron-hole dephasing
during transmission through long junctions
and during Andreev reflections given by matrices $U$.
In the mapping, dephasing is a source of phase gained
between the barriers in Fig.~\ref{potstructfig}.
In short constrictions, $L=0$, the resonant condition reads
$arg(a_{k})=m\pi$, where $m$ is integer, and it is fulfilled
near the gap edges, $E_{k}\approx\pm\Delta$.
(For energies outside the gap the phase gained is constant, $\pm\pi$,
but for $|E/\Delta|-1 > D^2$ the resonance vanishes because of the leakage.)
The resonance effectively removes two barriers so the bare
transmission probability is increased by the factor $D^{-2}$.
It is easy to see that at voltages $eV=2\Delta/n$ it is possible
to have simultaneously two resonances with corresponding indices
related as $k'=k+n$. In this case the transmission probability
$|t_{n0}|^2$ is enhanced by the factor $D^{-3}$ if the resonances are next
to each other ($n=1$), otherwise the enhancement factor is $D^{-4}$.

\begin{figure}
\centerline{\psfig{figure=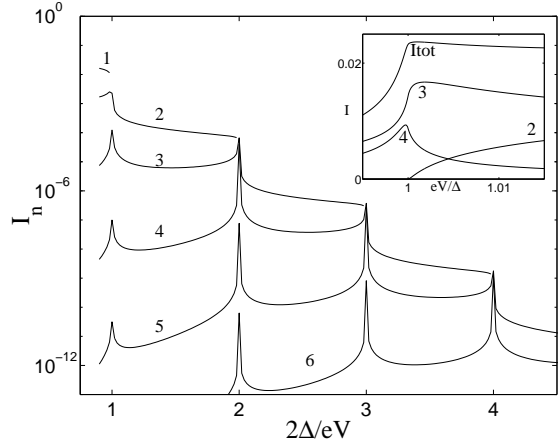,height=6cm}}
\caption{The SGS of a SIS junction with transparency
$D=0.01$ divided into different $n$-particle currents, for
$n\in\{1,6\}$. The inset shows the SGS of a SIS junction
with transparency $D=0.1$ close to $eV=\Delta$ and the contributions
from 2-, 3- and 4-particle currents.}
\label{SIS_SGS}
\end{figure}

Finally there are the resonances due to reflection from $\pm\infty$
in the denominator of Eq.~(\ref{Gdef}), boundary resonances.
The conditions for resonance are $(1-a_{0}^{2}r_{0-}r_{n0}^{+})=0$ and
$(1-a_{n}^{2}r_{n+}r_{n0}^{-})=0$, which is possible for 
$E=-\Delta$ or $E_{n}=\Delta$ respectively. These resonances
are only important when they overlap with resonances in the
bare transmission amplitude. An interplay of the resonances
in the bare transmission amplitude $t_{n0}$ and the boundary
resonances produces current peaks in the subharmonic gap structure
(Fig.~\ref{SIS_SGS}).

As an example of this interplay let us discuss the 4-particle
current in a short junction
\footnote{The more complex cases of junctions with
Breit-Wigner resonances and $d$-wave junctions are discussed in 
Refs.~\cite{Joh2} and \cite{Lofw} respectively.}, 
which possesses all typical features.
The 4-particle current ($I_{4}$) (see Fig.~\ref{SIS_SGS})
has an  onset at $2\Delta/eV=4$. In the voltage region
$3<2\Delta/eV<4$ there are no resonances in $t_{40}$ and
therefore the magnitude of the current is $I_{4}\propto D^4$.
For voltages $2\Delta/eV<3$ a single resonance in $t_{40}$ becomes
possible, which gives an enhancement of the current to 
$I_{4} \propto D^4 |ln(D)|$. Close to $2\Delta/eV=3$ this resonance
overlaps with a boundary resonance which
results in a current peak with the magnitude $(I_{4})_{max}\propto D^{3}$.
At $2\Delta/eV=2$ there is a double resonance, $E_{1}=-\Delta$ and
$E_{3}=\Delta$, which results in a current peak of order
$(I_{4})_{max}\propto D^{2}$. The voltage $2\Delta/eV=1$ is a special case because
the two resonances are next to each other so the enhancement of the
transmission probability is only $D^{-3}$, resulting in a current peak
$(I_{4})_{max}\propto D^{3}$. At this voltage the 4-particle current
is weakened by the leaky Andreev reflection outside the gap.

\section{Conclusions}
In conclusion, we have presented an approach, for analyzing the SGS in
quantum superconducting junctions, where MAR is
treated as a transport problem in energy space. Such an approach is natural
for the inelastic scattering problem represented by MAR. Using scattering
theory formalism, we have derived a
mapping of MAR onto a problem of transmission through a 1D wave guide with a
built in multiple tunnel barrier structure. 
There is no current conservation in this transmission problem due to a 
leakage outside the wave guide. The charge current in the
original superconducting junction is expressed through this leakage,
which is identical to the probability current of outgoing quasiparticles
in the superconductors. 
In terms of such a mapping, the SGS is explained
in terms of resonances in the transmission along the energy axis.
Three different types of resonances have been found which are important for
the interpretation of SGS:
(i) superconducting resonances induced by electron-hole dephasing and
the Andreev reflections, (ii) boundary resonances induced by reflections 
from the boundaries of the wave guide, and (iii) normal electron 
(Breit-Wigner) resonances, which may exist  
in addition to former two types of the resonances, which are always present
in MAR. The mapping allowed us to separate currents
associated with $n$-particle transmission processes
and to prove the cancellation of non-physical ground
state currents, which is equivalent to the Pauli exclusion principle.
\vspace*{-.2cm}

\end{document}